# Retinoblastoma Protein Is the Likely Common Effector for Distinct Anti-Aging Pathways

Razvan Tudor Radulescu

*Molecular Concepts Research (MCR), Munich, Germany*

E-mail: ratura@gmx.net

**ABSTRACT**

The multiple worlds of genetically manipulated laboratory organisms such as transgenic mice or worms with certain gene mutations are somewhat reminiscent of parallel worlds in quantum mechanics. So are various models of aging tested in such organisms. In this context, the tumor suppressor p53 has been found to either accelerate or delay aging, the latter, for instance, in conjunction with ARF, another tumor suppressor, as shown very recently. To more easily determine which of these artificial settings comes closest to real life, I discuss here their features in the light of my protein structure-based insights that have led me to propose a physiological anti-aging role for the retinoblastoma tumor suppressor protein (RB) over the past four years.

About two years ago, Serrano and coworkers reported that premalignant experimental tumors display features of senescence (1). The authors presented this induction of cellular aging as a defense mechanism mounted by the host in response to an oncogene-initiated neoplastic transformation (1). In this context, these investigators also qualified the tumor suppressor p53 as an "oncogene-induced senescence effector" (1). As such, this study supported the widely held view according to which the processes of cancer and aging exclude one another or yet that the sense of aging is to oppose the occurrence of neoplasias (2).

This month, however, Serrano *et al.* (3) communicated results that contradict the alleged dichotomy between senescence and cancer, yet they failed to quote their own paper of 2005 and explained the obvious data contradiction by different experimental conditions. Specifically, these investigators attributed those previous cases in which p53 had previously been found to accelerate aging to its "permanent activation" (3)- due to "truncation of p53 domains or constitutive endogenous damage" (3)- whereas their own results of this month led them to assume that increasing the activity of ("normally regulated") p53 produces delayed aging (3). Moreover, they showed data implying that p53 cooperates with ARF, which is another important tumor suppressor, in delaying aging (3).

Yet, these interpretations *per se* cannot solve the underlying puzzle as to which of the discussed observations on p53 is ultimately correct in terms of reflecting normal physiology since all of them are based on data obtained with genetically manipulated and thus abnormal mice. It follows that, by definition, p53 or any other gene product cannot be "normally regulated" in such animals. Nevertheless, some of these models may represent an useful approximation of biological events in real life, but in order to determine which of these models complies with reality one needs to take into account additional considerations.

In 2003, I derived a novel concept on a dual inhibition of cancer and aging from my identification of a sequence in the aminoterminus of retinoblastoma protein (RB) predicting that this tumor suppressor may antagonize the complex formation between insulin and its receptor (4). Since Kenyon and coworkers had found that insulin´s activation of its receptor may contribute to accelerating aging (5), I concluded that, conversely, an interference of RB with the physical interaction between insulin and its receptor may counteract, besides neoplastic transformation, also the process of senescence in entire organisms (4).

Subsequently, I detected similarities between the amino acid sequences of RB and the anti-aging protein Klotho, thus reinforcing my concept on an anti-aging role of RB and also assigning a possible tumor suppressor role to Klotho (6).

Interestingly, these previous publications of mine suggesting a connection between the functions of tumor suppression and anti-aging were echoed in part by the 2006 report by Kenyon and coworkers who demonstrated that worms with certain insulin receptor gene mutations display both an increased life span and an augmented resistance towards tumor formation (7).

Predictably, however, this worm model cannot entirely reflect equivalent phenomena in higher eukaryotes since the latter organisms have developed compensatory mechanisms to circumvent aberrancies in insulin signal transduction arising from a dysfunction of the insulin receptor. Based on this knowledge, an alternative model of aging involving an hyperinsulinemia secondary to insulin receptor defects and the subsequent interaction of insulin with retinoblastoma protein in the cell nucleus was presented (8).

I then tied together my own perception of aging (8) with the Kenyon concept on senescence to propose in an unifying framework that, in both aging models, retinoblastoma protein is being inactivated: in the first case, due to the binding of insulin to RB and, in the second case, through RB hyper-phosphorylation following the phosphorylation cascade elicited by insulin receptor activation (9).

This framework should help establish the roles of p53 and ARF in natural senescence. Given that it is known that RB acts as a downstream effector for both p53 (10,11) and ARF (12) to inhibit cell cycle progression and, as a result, to prevent cancer, it is conceivable that the anti-aging effects of p53 and ARF now observed in experimental models (1) have been mediated by RB, thus lending further support to the notion that RB is not only the key checkpoint in the control of cancer, but also of aging (9). Since cancer occurs predominantly in older individuals, my present molecular insight should significantly contribute to implementing the clinical rationale of reducing the incidence of malignancies by promoting longevity.